\language1
\magnification=1300
\vsize=23truecm
\hsize=15.6truecm
\tolerance=10000

\def\head #1 #2 { \headline={\vbox{ \line{\strut \TeX\hss #1--#2} \hrule}\hss} }
\def\ind {\noindent}
\def\tini {\mathop{\longrightarrow}}
\font\myfo=cmr8

\footline{\hss\rm\quad\hss}
\centerline{\bf GENERALIZED MOYAL STRUCTURES IN PHASE-SPACE,}
\centerline{\bf KINETIC EQUATIONS AND THEIR CLASSICAL LIMIT}
\centerline{\bf II: APPLICATIONS TO HARMONIC OSCILLATOR MODELS}
\vskip 0.8truecm
\centerline{\myfo CONSTANTINOS TZANAKIS  }
\centerline{\myfo UNIVERSITY OF CRETE }
\centerline{\myfo 74100 RETHYMNON, CRETE, GREECE }
\vskip 0.2truecm
\centerline{\myfo ALKIS P. GRECOS\footnote{$^*$}{\myfo Association Euratom, Etat
Belge}}
\centerline{\myfo Service de Physique statistique et plasmas}
\centerline{\myfo UNIVERSIT\'E LIBRE DE BRUXELLES}
\centerline{\myfo CAMPUS PLAINE (CP 231)}
\centerline{\myfo 1050 BRUSSELS, BELGIUM}
\vskip 0.2truecm
\centerline{\myfo POLYXENI HATJIMANOLAKI}
\centerline{\myfo INTERNATIONAL BACCALAUREATE,} 
\centerline{\myfo KOSTEAS-GITONAS SCHOOL,}
\centerline{\myfo ATHENS 10534, GREECE }
\vskip 1truecm
\ind ABSTRACT {\myfo
The formalism of generalized Wigner transformations developped in a previous
paper, is applied to kinetic equations of the Lindblad type for quantum 
harmonic oscillator models. It is first applied to an oscillator coupled to
an equilibrium chain of other oscillators having nearest-neighbour 
interactions. The kinetic equation is derived without using the so called 
rotating-wave approximation. Then it is shown that the classical limit of
the corresponding phase-space equation is independent of the ordering of
operators corresponding to the inverse of the generalized Wigner transformation,
provided the latter is involutive. Moreover, this limit equation, which
conserves the probabilistic nature of the distribution function and obeys an
H-theorem, coincides with the kinetic equation for the corresponding classical
system, which is derived independently and is distinct from that usually
obtained in the litterature and not sharing the above properties. Finally
the same formalism is applied to more general model equations used in quantum
optics and it is shown that the above results remain unaltered.}
\vfill{\eject}
\leftline{\bf 1. INTRODUCTION}
\bigskip
\par
In a previous paper ([1], here after called paper I), we studied the properties
of generalized Weyl and Wigner transformations of quantum operators and 
classical phase-space functions respectively and with their help we presented
a phase-space formulation of quantum evolution equations. In the case of open
systems interacting with a large equilibrium bath we gave an explicit 
phase-space representation of quantum kinetic equations of the Lindblad type.
As it is well-known, such equations are satisfactory in the sense that they
conserve the probabilistic interpretation of the density matrix and obey an
H-theorem ([2], [3], [4] section V, [5] proposition 4.3, [6] section 4.3.).
They
are obtained in the context of various approaches to kinetic theory that employ
some more or less systematic approximation scheme to the exact dynamics of the
system ([3]-[5], [7]-[12]). In the present paper we illustrate the formalism
of paper I by applying it to kinetic equations of this kind, for particular 
harmonic oscillator models.
\par
In section 2, we first apply the general kinetic equation for an open system
weakly interacting with a bath at equilibrium, that follows in the context of
the formal theory presented in [5],
a brief account of which is given in Appendix 3, 
to the case of a {\it classical} harmonic
oscillator interacting with a (linear) harmonic chain, assuming nearest
neighbour interactions. 
Moreover, the chain is
considered to be at canonical equilibrium. We show that in the thermodynamic
limit of an infinite chain, the interaction Hamiltonian takes a simple separable
form. The derivation uses functional methods and leads to an equation of the
Fokker-Planck type in the {\it phase-space} of the oscillator.
\par
Then in section 3 the corresponding {\it quantum} system is treated in the 
context of the same formalism, leading to a kinetic equation well-known in 
the litterature but either introduced phenomenologically or derived with the aid
of additional approximations. Its phase-space representation via an arbitrary
generalized Wigner transformation is accordingly given. It is also shown that
{\it its classical limit} is {\it unique}, as long as the corresponding
generalized Moyal bracket is a deformation of the Poisson bracket, and
{\it coincides with the classical equation obtained in section 2}, which
for the sake of completeness was derived there independently, thus
exhibiting the consistency of our formalism, at least in this special
case.
\par
Finally in section 4, a more general model equation of the Lindblad type for
a harmonic oscillator linearly coupled to a bath of other oscillators, that
has been considered in the litterature especially in quantum optics
([13], [14] and references there in) is phase-space transformed with the aid 
of the formalism of paper I. This generalizes previous results of [14]
and easily implies that once again {\it its classical limit is unique} under
the condition stated above.
\vskip 2truecm
\leftline {\bf 2. CLASSICAL OSCILLATOR LINEARLY COUPLED TO A}
\leftline {\bf \hskip 0.65truecm HARMONIC CHAIN}
\bigskip
\par
In this section we will study a classical harmonic oscillator model and in
the next section our results will be related to those of the corresponding 
quantum system. The model consists of a harmonic oscillator (the subsystem
$\Sigma$) coupled to an {\it infinite} set of harmonic oscillators 
(the reservoir $R$), at canonical equilibrium. It is assumed that the 
interaction is (bi)linear in the coordinates of $\Sigma$ and $R$. The
Hamiltonian reads
$$\hskip 2truecm H=H_{\Sigma}+H_R+\lambda 'H_I \hskip 8.5truecm (2.1)$$
$$\hskip 2truecm H_{\Sigma}={p'^2\over 2M}+{1\over 2}kq'^2\hskip 8.7truecm (2.2a)$$
$$\hskip 2truecm H_R = {1\over 2m} \sum_k p_k'{^2} + {1\over 2} \sum_{k,l} q_k'h_{k-l}'q_l'
\hskip 5.7truecm (2.2b)$$
$$\hskip 2truecm H_I=\sum_k \epsilon_k'q_k'q' \hskip 9.3truecm (2.2c)$$
where $\lambda'$ is a coupling parameter and the remaining symbols have  an
obvious meaning. Performing the (canonical) scale transformation
$$p=M^{-{ 1\over 2}}p'\ \ \ \ \ q=M^{1\over 2}q'$$
$$p_k=m^{-{1\over 2}}p_k'\ \ \ \ \  q_k=m^{1\over 2}q_k'$$
we set 
$$\hskip 2truecm H=H_{\Sigma}+H_R+\lambda H_I,\ \ \ \ \ \lambda = \sqrt{m\over M} \lambda'\hskip 5truecm (2.3)$$
$$\hskip 2truecm H_{\Sigma}= {p^2\over 2}+\Omega_0^2q^2,\ \ \ \ \ \Omega_0^2={k\over M}\hskip 6.5truecm (2.4a)$$
$$\hskip 2truecm H_R={1\over 2}\sum_k p_k^2+{1\over 2}\sum_{k,l} q_kh_{k-l}q_l,\ \ \ \ \ h_k={h_k'\over m} \hskip 3.7truecm (2.4b)$$
$$\hskip 2truecm H_I=\sum_k\epsilon_k q_k q\ \ \ \ \ \epsilon_k = {\epsilon_k'\over m} \hskip 7.3truecm (2.4c)$$
Actually in defining $H_R$ we may take any positive-definite
symmetric matrix $h_{kl}$ and not only a codiagonal one. This is readily seen
for a {\it finite} chain by taking an orthogonal matrix $\gamma_{mn}$
diagonalizing $h_{mn}$ and making the (canonical) transformation
$$(q_k',p_k') \rightarrow (\tilde q_k,\tilde p_k)$$
$$\tilde p_m=\sum_k\gamma_{km}p_k',\ \ \ \ \ \tilde q_m=\sum_k\gamma_{mk}q_k'$$
However for an {\it infinite} system the diagonalization procedure is more
involved since subtleties appear due to the (presummably simultaneous) 
existence of a continuous and a point spectrum of $h_{kl}$. 
Nevertheless, if it posseses a complete set of eigenvectors, normal coordinates
can be defined in analogy with (2.6) below, hence (2.8) follows and consequently
kinetic equation ((2.26) below) remains unmodified.
In view of the above discussion it is reasonable to assume in the present
case
$$\hskip 2truecm h_k=h_k^*=h_{-k} \geq 0 \hskip 8.8truecm (2.5)$$
where $h_k^*$ denotes the complex conjugate.
\par 
This model has been used extensively,
particularly in the case of nearest neighbour interactions the various
calculations being performed for a {\it finite} chain and then taking,
the thermodynamic limit
(see e.g. [15] eq(1), [16] eq(18), [17]-[19] and indirectly [20]).
It should be noted that one often considers
$\sqrt{{m\over M}}$ as a small parameter (Brownian motion proper)  and not
just the coupling parameter $\lambda'$ as in this paper.
\par
In case of an infinite number of nonvanishing $h_k$'s it will be assumed that 
$h_k \rightarrow 0$ as $\mid k\mid \rightarrow +\infty$ sufficiently rapidly
and as a matter of fact, they will be defined as the coefficients of a periodic
function. Normal coordinates are then introduced by the Fourier series
$$\hskip 1truecm \phi(\theta) = {1\over \sqrt{2\pi}} \sum_{k=-\infty}^{+\infty}
q_k e^{ik\theta}\ \ \Leftrightarrow\ \ q_k = {1\over \sqrt{2\pi}}\int_{-\pi}^{\pi}
d\theta \phi(\theta) e^{-ik\theta} \hskip 2.08truecm (2.6a)$$
$$\hskip 1truecm \pi(\theta)={1\over \sqrt{2\pi}} \sum_{k=-\infty}^{+\infty} p_k e^{-ik\theta}
\ \ \Leftrightarrow\ \ p_k= {1\over \sqrt{2\pi}}\int_{-\pi}^{\pi} d\theta \pi(\theta)
e^{ik\theta} \hskip 2truecm (2.6b)$$
This transformation is canonical and $\xi_k(\theta)\equiv{1\over \sqrt{2\pi}}e^{ik\theta}$
may be considered as a complete orthonormal set of vectors
$$\hskip 1truecm \sum_{k,l} \xi_k^*(\theta)\xi_l(\theta') = \delta(\theta-\theta') \hskip 8.4truecm (2.7a)$$
$$\hskip 1truecm \int_{-\pi}^{\pi} d\theta \xi_k(\theta)\xi_l^*(\theta) = \delta_{k,l} \hskip 8.7truecm (2.7b)$$
where $\delta(\theta)$ is periodic with period $2\pi$ and $\delta_{k,l}$ is the
Kronecker delta. Introducing (2.6) in (2.4) we get
$$\hskip 1truecm H_R={1\over 2}\int_{-\pi}^{\pi}d\theta(|\pi(\theta)|^2+\omega^2(\theta)
|\phi(\theta)|^2) \hskip 5.8truecm (2.8a)$$
$$\hskip 1truecm H_I = \int_{-\pi}^{\pi}d\theta u^*(\theta)\phi(\theta) q \hskip 8.7truecm (2.8b)$$
$$\hskip 1truecm u(\theta)={1\over\sqrt{2\pi}}\sum_{n=-\infty}^{+\infty}\epsilon_n
e^{in\theta}\ \ \Leftrightarrow\ \ \epsilon_n={1\over \sqrt{2\pi}}\int_{-\pi}^{\pi}
d\theta u(\theta)e^{-in\theta} \hskip 2truecm (2.9)$$
$$\hskip 1truecm \omega^2(\theta)\equiv\sum_{k=-\infty}^{+\infty}h_ke^{ik\theta}
=\omega^2(-\theta) \geq 0 \hskip 6.4truecm (2.10)$$
because of (2.5). Note that $\epsilon_n$ is real and that it is reasonable to
consider that $\epsilon_n=\epsilon_{-n}$ (cf. e.g. [15] eq.(10) for the finite
case), so that $u(\theta)$ is real and even.
\par
As we have used the basis $\{{1\over \sqrt{2\pi}}e^{ik\theta}\}$, the variables
$\phi(\theta), \pi(\theta)$ are complex-valued functions of $\theta$. Had we
chosen the basis $\{{1\over \sqrt{2\pi}},{1\over \sqrt{\pi}}\cos k\theta,
{1\over \sqrt{\pi}}\sin k\theta,k\in {\cal N}\}$, the corresponding variables would
be real. Note also that by (2.9), only the real part $\Re\phi(\theta)$ contributes
to $H_I$ in (2.8b).
\par
In view of the above discussion, we will consider in the rest of this section 
a simple version of the
Hamiltonian (2.8) namely
$$\hskip 1truecm H_R={1\over 2}\int d\omega (\pi^2(\omega)+\omega^2\phi^2(\omega))\hskip 6.8truecm (2.11a)$$
$$\hskip 1truecm H_I=\int d\omega u(\omega) \phi(\omega) q\equiv Wq \hskip 7.5truecm (2.11b)$$
Where $\pi(\omega),\phi(\omega)$ are real, {\it the intergration interval is
symmetric about the origin and $u(\omega)$ is real and even} (this is not an
essential assumption; see end of Appendix 1). 
It will become evident that results for the general case (2.8) 
follow by replacing $u(\omega)$ with
$u^2(\omega){d\theta\over d\omega}$ in the final expressions (see also the 
discussion in Appendix 1).
\par
Clearly Hamilton's equations for the total system are
$$\hskip 1truecm \dot q(t)=p(t) \hskip 11truecm (2.12a)$$
$$\hskip 1truecm \dot p(t)=-\Omega^2_0q-\lambda\int d\omega u(\omega) \phi(\omega) \hskip 6.8truecm (2.12b)$$
$$\hskip 1truecm \dot \phi(\omega)={\delta H\over\delta \pi(\omega)}=\pi(\omega) \hskip 8.8truecm (2.12c)$$
$$\hskip 1truecm \dot \pi(\omega)=-{\delta H\over\delta\phi(\omega)}=-\omega^2\phi(\omega)-
\lambda u(\omega)q\hskip 5.8truecm (2.12d)$$
using functional derivatives. Because the Hamiltonian is quadratic and the 
reservoir in a canonical equilibrium state)
functional methods (see e.g. [24],[25]) will be used in the following to treat
ab initio the infinite system.
\par
Because of its linearity, the model is exactly solvable, the solution being 
obtained by solving the generalized eigenvalue problem
$$\Omega_0^2Z_{\nu}+\lambda\int d\omega'u(\omega')\xi_{\nu}(\omega)=
\nu Z_\nu$$
$$\lambda u(\omega) Z_\nu + \omega^2 \xi_\nu(\omega)=\nu\xi_\nu(\omega)$$
For $\nu$ outside the interval $\omega^2$ varies, we get
$$\xi_\nu(\omega)=-\lambda{u(\omega)\over \omega^2-\nu}Z_\nu$$
$$\biggl(\Omega_0^2-\nu-\lambda^2\int d\omega {u^2(\omega)\over\omega^2-\nu}\biggr)
 Z_\nu=0$$
Thus for nontrivial solutions
$$\Omega_0^2-\nu=\lambda^2\int d\omega {u^2(\omega)\over\omega^2-\nu}$$
Nonnegativity of the Hamiltonian excludes negative values for $\nu$ which lead 
to instabilities, hence we must impose the constraint (cf. [15] eq.(24))
$$\hskip 5truecm \Omega_0^2\geq \lambda^2\int d\omega {u^2(\omega)\over \omega^2} \hskip 5.2truecm (2.13)$$
which inter alia implies that at the origin $u^2(\omega)$ is $o(\omega^2)$ and if the
range of $\omega$ is infinite $u^2(\omega)$ cannot grow at infinity more rapidly than
$\mid\omega\mid^{2-\epsilon}$ for $0<\epsilon <1$. Clearly these considirations
imply that $\Omega_0$ cannot vanish, otherwise the model will necessarily lead to
instabilities.
Note however that, had we taken $H_I=\sum_ku_k(q_k-q)^2 $ the problem would
not arise as $\Omega_0^2$ should be replaced by $\Omega_0^2+\lambda\sum_ku_k$
\par
Although the model is solvable in principle, we will consider weak coupling of
$\Sigma$ with $R$, and we will obtain the kinetic equation for the oscillator by 
employing the general formalism of [5] that was briefly discussed in section 4 of 
paper I and Appendix 3 here,that is, compute the generator $\Phi$ of Appendix
3. Because of the separability of $H_I$, eq.(2.11b), we can use the 
corresponding general formulas of [5] eqs (4.22), $(4.22')$. To this end we need
$$\hskip 1truecm q(t)\equiv e^{iL_\Sigma t}q=q\cos\Omega_0t+{p\over \Omega_0}
\sin\Omega_0t \hskip 5.8truecm (2.14a)$$
$$W(t)\equiv e^{iL_R t}W\equiv \int d\omega\ u(\omega) \phi(\omega,t)=\hskip 6truecm $$
$$\hskip 4.5truecm \int d\omega\ u(\omega) (\phi(\omega)\cos\omega t+{\pi(\omega)\over \omega}
\sin\omega t) \hskip 2.5truecm (2.14b)$$
where we write $L_\Sigma=i\{H_\Sigma,\cdot\}$ etc for the various Liouville
operators corresponding to (2.1) and we used that for $\lambda=0$, (2.12) gives
the solution $q(t), \phi(\omega,t)$. In the present case eqs (4.22), $(4.22')$ of
[5] for the probability distribution $f$ of the harmonic oscillator are 
(cf. (2.11b)):
$${\partial f\over\partial t}=\{H_\Sigma-\lambda^2F,f\}+\hskip 8truecm $$
$$\hskip 1truecm +{\lambda^2\over 2}\sum_n(\tilde h(\omega_n)\{(P_nq)^*,\{P_nq,f\}\}+
\tilde g(\omega_n)\{(P_nq)^*,(P_nq)f\})\hskip 1.7truecm  (2.15)$$
$$\hskip 1truecm F={1\over 2} \sum_n(\bar h(\omega_n)\{(P_nq)^*,P_nq\}+\bar g(\omega_n)(P_nq)^*
P_nq) \hskip 3.5truecm (2.16)$$
Here $P_n,\omega_n$ are the eigenprojections and eigenvalues of $L_\Sigma$,
$$\hskip 1truecm \tilde h(\omega)=\int_{-\infty}^{+\infty}ds\ e^{i\omega s}h(s),\quad\ 
h(s)\equiv <W\ W(s)> \hskip 3.5truecm (2.17a) $$
$$\hskip 1truecm \tilde g(\omega)=\int_{-\infty}^{+\infty}ds\ e^{i\omega s}g(s),\quad\ 
g(s)\equiv <{W\ W(s)}> \hskip 3.5truecm (2.17b) $$
$$\hskip 1truecm \bar h(\omega)=\int_0^{+\infty}ds\ e^{i\omega s} h(s)\hskip 8.3truecm (2.17c)$$
$$\hskip 1truecm \bar g(\omega)=\int_0^{+\infty}ds\ e^{i\omega s} g(s)\hskip 8.3truecm  (2.17d)$$
and $<\ \ \ \ >$ denotes the average over the chain, assumed in a canonical
equilibrium state
$$\hskip 1truecm \rho_R={1\over Z}e^{-{\beta\over 2}\int d\omega(\pi^2(\omega)+\omega^2\phi^2
(\omega)))}={1\over Z}e^{-\beta H_R} \hskip 4.5truecm (2.18a)$$
$$\hskip 1truecm Z=\int\delta\phi\delta\pi\ e^{-\beta H_R} \hskip 9.3truecm (2.18b)$$
clearly
$$\omega_n=n\Omega_0,\quad\ n\in {\cal Z}$$
$$\hskip 1truecm P_nA(\theta,\zeta)={1\over 2\pi}\int_{-\pi}^\pi d\theta' e^{in(\theta-\theta')}
A(\theta',\zeta) \hskip 5.5truecm (2.19) $$
with $\zeta,\theta$ the action-angle variables of the oscillator:
$$\hskip 1truecm q=\biggl({2\zeta\over\Omega_0}\biggr)^{1\over 2}\sin\theta,\ \ \ \ \ 
{p\over\Omega_0}=\biggl({2\zeta\over\Omega_0}\biggr)^{1\over 2}\cos\theta \hskip 5.2truecm (2.20)$$
A simple calculation using (2.19), (2.20) yields
$$\hskip 1truecm P_nq={1\over 2}(q-{ip\over \Omega_0})\delta_{n,1}+{1\over 2}(q+{ip\over\Omega_0}) \delta_{n,-1} \hskip 5.3truecm (2.21)$$
On the other hand explicit calculation of (2.17) involves certain simple functional
integrals and derivatives. The details are given in Appendix 1. The results are:
$$\hskip 1truecm \tilde h(\omega)={2\pi u^2(\omega)\over \beta\omega^2},\quad\ 
\tilde g(\omega)=i\beta\omega\tilde h(\omega) \hskip 6.2truecm (2.22)$$
$$\hskip 1truecm \bar h(\omega)={\tilde h(\omega)\over 2}+i\int d\omega' {u^2(\omega')\over
\beta \omega'}{\omega\over \omega^2-\omega'^2}\hskip 5.7truecm  (2.23a)$$
$$\hskip 1truecm \bar g(\omega)={\tilde g(\omega)\over 2}-\int d\omega' {u^2(\omega')\over
\omega^2-\omega'^2}\hskip 7.2truecm (2.23b)$$
Notice that because of (2.18), the second of (2.22) follows from the first and 
proposition 4.5 of [5]. An elementary calculation using (2.21), (2.22) gives
\smallskip
\ind dissipative part of (2.15)=
$$\hskip 1truecm ={\lambda^2\pi\over 2\beta}{u^2(\Omega_0)\over\Omega_0^2}\biggl[{1\over \Omega_0^2}
{\partial^2f\over \partial q^2}+{\partial^2f\over \partial p^2}+2\beta
({\partial\over \partial q}(qf)+{\partial\over \partial p}(pf))\biggr]\hskip 2.6truecm (2.24)$$
Similarly (2.21), (2.23) substituted in (2.16) give
$${1\over 2}\sum_n\bar h(\omega_n)\{(P_nq)^*,P_nq\}={1\over 2\beta}
\int d\omega {u^2(\omega)\over \omega^2}{1\over \Omega^2_0-\omega^2}=constant$$
$${1\over 2}\sum_n\bar g(\omega_n)(P_nq)^*P_nq=-H_\Sigma{1\over 2\Omega_0^2}
\int d\omega {u^2(\omega)\over \Omega_0^2-\omega^2} $$
Since the first does not contribute to the first term on the r.h.s. of (2.15)
we may write (2.16) as:
$$\hskip 1truecm F={\Delta(\Omega_0)\over \Omega_0}H_\Sigma ,\ \ \ \ \ 
\Delta(\Omega_0)={1\over 2\Omega_0}\int d\omega {u^2(\omega)\over\omega(\omega-
\Omega_0)} \hskip 3.5truecm (2.25)$$
where we used that the range of $\omega$ is symmetric about the origin.By (2.24),
(2.25), equation (2.15) takes the form
$${\partial f\over \partial t}=\biggl(1-{\lambda^2\Delta(\Omega_0)\over\Omega_0}\biggr)
\{H_\Sigma,f\}+\hskip 9truecm $$
$$\hskip 1truecm +{\lambda^2\pi\over 2\beta}{u^2(\Omega_0)\over \Omega_0^2}
\biggl({\partial\over\partial q}\Bigl({1\over \Omega_0^2}
{\partial f\over\partial q}+2\beta q f\Bigr)+{\partial\over\partial p}\Bigl({\partial f\over \partial p}+2\beta p f\Bigr)\biggr)
\hskip 2.7truecm (2.26) $$
we may notice that as expected, the Maxwell-Boltzman (MB) distribution 
$e^{-\beta H_\Sigma}$ is a stationary solution of (2.26) and $F$ is an integral
of the unperturbed motion, results that are special cases of proposition 4.1, 4.2
of the general formalism of [5] that led to (2.15).
\par
It may also be remarked that in the litterature, a different kinetic equation has 
been derived for this model ([21] eq.(20) in connection with eq(31) of [15])
$${\partial f\over \partial t}-(1-{\lambda^2\Delta(\Omega_0)\over\Omega_0})
\Omega_0^2 q{\partial f\over \partial p}+p{\partial f\over\partial q}=\hskip 8truecm $$
$$\lambda^2{\partial\over \partial p}\biggl({\pi\over \beta}{u^2(\Omega_0)\over\Omega_0^2}
({\partial f\over\partial p}+2\beta p f)+{\chi(\Omega_0)\over\Omega_0}{\partial f
\over \partial q}\biggr)$$
$$\chi(\Omega_0)=\int d\omega {u^2(\omega)\over\omega^2(\omega-\Omega_0)}$$
In fact it can be shown that this equation is the second order term in the
$\lambda$-expansion of the so-called Generalized Master equation (GME), that 
follows by projecting the state of the total system $\Sigma+R$ to that of $\Sigma$,
namely eqs. $(3.7')$ or (5.1) of [5] (see also Appendix 3). Clearly this equation has not a 
nonegative-definite 2nd order coefficient matrix and therefore it does not conserve
the positivity of $f$ and for that matter obeys no H-theorem. Moreover it does not 
have the MB distribution $ce^{-\beta H_\Sigma}$ as an equilibrium solution (cf. [5]
sections 4, 5, [23] section 4, 5). A somewhat different equation but still with the 
structure of (2.27), hence the same undesirable features, has been obtained by a similar
analysis of the GME, ([30],[31]; see also [34] and the discussion in [5] section
[5]).
Sometimes the mixed derivatives-term
is neglected arguing that it is "small" ([21] p.60) so that a conventional 
Fokker-Planck equation results. However even in this case the MB distribution is
not stationary, that is even for a chain at canonical equilibrium, the external
oscillator does not evolve toward such an equilibrium state! Moreover it should
be noticed that in this case this neglect leads to different coefficients for 
the conventional part of the Fokker-Plank equation (compare (2.26), (2.27)). This
is a general feature that can be understood on the basis of the general 
formalism presented in [5], since the markovian approximation of the GME leads 
to an equation ($(3.7')$ of [5] for the generator $\Theta$ in Appendix 3)
totally different from that obtained by our
formalism ($(3.11')$ of [5] for the generator $\Phi$ in Appendix 3). This 
fact has been verified in other models as
well (e.g. compare eqs (4.1), (4.2) with (4.4), (4.5) of [33] section 4).
A further argument
in favor of (2.26) is provided in the next section, where we
show that (2.26) is the unique classical limit of the kinetic equation of the
Lindblad type for the corresponding quantum system, which is well-known and widely
used, e.g. in quantum optics.
\vskip 2truecm
\leftline{\bf 3. KINETIC EQUATION FOR A QUANTUM HARMONIC }
\leftline{\bf \hskip 0.6truecm OSCILLATOR, WEAKLY-COUPLED TO A HARMONIC}
\leftline{\bf \hskip 0.6truecm CHAIN}
\bigskip
In this section we consider the {\it quantum} system corresponding to the
classical model of the previous section. It is easily seen that the Hamiltonian
has the form (hated quantities denoting operators-cf. paper I section 2 for the
notation):
$$\hskip 1truecm \hat H=\hat H_\Sigma+\hat H_R+\lambda\hat H_I=\hskip 9truecm $$
$$\hskip 1.8truecm =\hbar\Omega_0\hat a^+\hat a+\sum_k\hbar\omega_k\hat a^+_k\hat a_k+\lambda
\sum_k\hbar (\epsilon_k\hat a^+_k+\epsilon_k^*\hat a_k)(\hat a+\hat a^+)\hskip 2truecm (3.1)$$
where
$$\hskip 2truecm \hat a={1\over \sqrt{2\hbar}}(\sqrt\Omega_0\hat q+i{\hat p \over \sqrt\Omega_0})\ \ \ \ \ \ \ \ 
\hat a^+={1\over \sqrt{2\hbar}}(\sqrt\Omega_0\hat q-i{\hat p \over \sqrt\Omega_0})\hskip 1.8truecm (3.2)$$
are the creation-annihilation operators for the oscillator and $\hat a^+_k,
\hat a_k$ the corresponding quantities for the $k$-th bath oscillator, defined
similarly\footnote{$^{(1)}$}{\myfo Non-unit masses can be absorbed into $\hat q,
\hat p$ by the same rescaling we did for the classical system.}, and
$\Omega_0,\ \omega_k$ {\it are nonegative}.
\par
In fact it is readily seen that any generalized Wigner transformation sending
functions of $\hat q$ or $\hat p$ to functions of $q$ or $p$ respectively
(i.e. (3.17) of paper I holds) maps (3.1) to (2.4)(see the comments following it),
provided $\epsilon_k$ is real. For an infinite chain, in the thermodynamic
limit we make the replacements
$$\hskip 2truecm \omega_k\tini \omega\ \ \ \ \  
\epsilon_k\tini \epsilon(\omega)\ \ \ \ \  
\sum_k \tini \int d\omega\sigma(\omega)\hskip 4truecm $$ 
$\sigma(\omega)$ being the spectral density of the chain. Consequently, if 
$\sigma=1$, i.e. $\epsilon(k)=\epsilon(\omega_k)$, then (3.1) is mapped to
(2.11) with $u^2(\omega)\leftrightarrow 4\epsilon^2(\omega)\omega\Omega_0$.
A detailed comparison of the results of the two sections obviously requires
$\omega >0$ in the previous section and is discussed at the end of Appendix 1.
\par
In the litterature, especially in quantum optics, the Hamiltonian (3.1) 
is often treated in the so-called rotating-wave approximation where the terms 
$\hat a^+_k\hat a^+$, $\hat a_k\hat a$ 
are neglected (see e.g. [8] p. 336 and eq. (6.2.39), [9] ch. 3 section 3b,
implicity in [12] ch. 12, [22], [36] section 5.14). Sometimes
such an approximation is made at the end of the derivation of the kinetic 
equation ([11] p. 743 eq. (3.1)). To derive a Markovian master equation for
weakly coupled systems, such an approximation is not necessary and for our
purpose not desirable, since the Hamiltonian should reduce to the classical
one by a generalized Wigner transformation (see also Appendix 3). We indicate
here the steps to apply the formalism of [5] that is, compute the generator
$\Phi$ in Appendix 3, noting that $\hat H_\Sigma$ is
separable (cf. [35] for rigorous estimates of the approach to equilibrium). 
Specifically we apply (5.8) of paper I to (3.1), with
$$\hskip 1truecm \hat H_I=\hat W\hat q=\sum_k \hat W_k\hat q=\sum_k\hbar
(\epsilon_k\hat a_k^++\epsilon_k^*\hat a_k^+)(\hat a+\hat a^+)\hskip 3.5truecm (3.3)$$
if $\nu =n \omega_0,{\cal F}_n$ are the eigenvalues and eigenprojections
of ${1\over \hbar}[\hat H_\Sigma ,\cdot ]=L_\Sigma$ , $n\in {\cal Z}$,
then clearly
$$\hskip 2truecm {\cal F}_n\hat q=\sum_{m\in {\cal Z}}P_m\hat qP_{m-n}\hskip 9truecm $$
where $P_n$ are the eigenprojections of the number operator $\hat N=\hat a^+\hat a$. Using the well-known expression for its eigenvectors
$$\hskip 2truecm \hat N\vert n>=n\vert n>\hskip 9.3truecm (3.4a)$$
$$\hskip 2truecm \hat a\vert n>=\sqrt n\ \vert n-1>\hskip 1truecm \hat a^+\vert n>=\sqrt{n+1}\ \vert n+1>\hskip 2truecm (3.4b)$$
a simple calculation gives
$$\hskip 2truecm {\cal F}_n\hat q=(\hat a\delta_{n,-1}+\hat a^+\delta_{n,1})\hskip 7.8truecm (3.5)$$
On the other hand, similar expressions to (3.4) for the bath oscillator and
$$\hskip 2truecm e^{iL_Rs}\hat A=e^{i{\hat H_R\over \hbar}s}\hat Ae^{-i{\hat H_R\over \hbar}s}\quad ,\ L_R={1\over \hbar}[\hat H_R,\cdot ]\hskip 4.8truecm $$
give
$$\hskip 2truecm e^{iL_Rs}\hat a_k=e^{-i\omega_ks}\hat a_k
\quad\quad\quad e^{iL_Rs}\hat a^+_k=e^{i\omega_ks}\hat a^+_k\hskip 3.8truecm (3.6)$$
Therefore by (3.3)
$$\hskip 2truecm \hat W_k(s)\equiv e^{iL_Rs}\hat W_k=\hbar(\epsilon_ke^{i\omega_ks}\hat a^+_k+\epsilon_k^*e^{-i\omega_ks}\hat a^+_k)\hskip 3.2truecm (3.7)$$
As in the previous section, we assume the bath to be in canonical equilibrium
$$\hskip 2truecm \hat \rho_R={e^{-\beta\hat H_R}\over Tr(e^{-\beta\hat H_R})}\hskip 9.18truecm (3.8)$$
A simple calculation using (3.1) gives 
$$\hskip 2truecm Tr(e^{-\beta\hat H_R})=\prod_kTr(e^{-\beta\hbar \omega_k\hat a^+_k\hat a_k})=\prod_k{(1-e^{-\beta h\omega_k})}^{-1}\hskip 2truecm $$
$$\hskip 2truecm <\hat a_k\hat a_{k'}>=<\hat a^+_k\hat a^+_{k'}>=0\hskip 7.2truecm (3.9a)$$
$$\hskip 2truecm <\hat a^+_k\hat a_{k'}>={\delta_{k,k'}\over {e^{\beta\hbar\omega_k}-1}}\equiv {n_k\over \hbar}\delta_{k,k'}\hskip 5.8truecm (3.9b)$$
$$\hskip 2truecm <\hat a_{k'}\hat a^+_k>=\biggl({n_k\over \hbar}+1\biggr)\delta_{k,k'}\hskip 7.1truecm (3.9c)$$
where $<\hat A>=Tr(\hat \rho_R\hat A)$ and the $\hbar$-dependence in (3.9) is 
shown explicitly so that $n_k$ has a finite value $(\beta\omega_k)^{-1}$ in the 
classical limit. Using (3.9) we readily find
$$\hskip 2truecm h(s)\equiv <\hat W^+\hat W(s)>=\hbar^2\sum_k{\vert \epsilon_k\vert}^2
\biggl(e^{i\omega_ks}\biggl({n_k\over \hbar}+1\biggr)+e^{-i\omega_ks}{n_k\over \hbar}\biggr)\hskip 0.7truecm $$
Using (A.1.3) we get 
$$\tilde h(\omega)\equiv \int\limits_{-\infty}^{+\infty}ds\ e^{i\omega s}h(s)=$$
$$\hskip 2truecm 2\pi\hbar^2\sum_k{\vert \epsilon_k\vert}^2\biggl(\biggl({n_k\over \hbar}+1\biggr)\delta(\omega +\omega_k)+{n_k\over \hbar}\delta(\omega -\omega_k)\biggr)\hskip 2truecm (3.10a)$$
$$s(\omega)\equiv \Im\int\limits_0^{+\infty}ds\ e^{i\omega s}h(s)=
\hbar^2\sum_k{\vert \epsilon_k\vert}^2\biggl(\biggl({n_k\over \hbar}+1\biggr){1\over {\omega+\omega_k}}+{n_k\over \hbar}{1\over {\omega-\omega_k}}\biggr)\hskip 0.2truecm (3.10b)$$
We are now ready to obtain the kinetic equation for the density matrix $\hat \rho$
of the oscillator, by applying the above results to the general kinetic equation
(5.8) of paper I, which for the present model reads:
$${\partial \hat \rho \over \partial t}=-{1\over \hbar}[\hat H_\Sigma,\hat \rho]
+{{i\lambda^2}\over \hbar^2}\biggl[\sum_ns(n\Omega_0){({\cal F}_n\hat q)}^+({\cal F}_n\hat q),\ \hat\rho\biggr]+\hskip 4.5truecm $$
$$\hskip 1.2truecm {\lambda^2\over {2\hbar^2}}\sum_n\tilde h(n\Omega_0)\biggl(\Bigl[({\cal F}_n\hat q)\hat\rho,\ {({\cal F}_n\hat q)}^+\Bigr]+\Bigl[{\cal F}_n\hat q,\ \hat\rho{({\cal F}_n\hat q)}^+\Bigr]\biggr)\hskip 2truecm (3.11)$$
In the thermodynamic limit of an infinite chain, applying (3.10), (3.5) to
(3.11) using that $\omega>0$ and making straightforward reductions we get
([26] ch. III)
$${\partial \hat \rho \over \partial t}=-{i\over \hbar}\Bigl(1-\lambda^2
{\Delta(\Omega_0)\over \Omega_0}\Bigr)[\hat H_\Sigma ,\hat \rho]+\hskip 8.7truecm $$
$$+\pi\lambda^2{\vert\epsilon(\Omega_0)\vert}^2\sigma(\Omega_0)
\biggl({n(\Omega_0)\over \hbar}\Bigl([\hat a^+\hat\rho,\hat a]+
[\hat a^+,\hat\rho\hat a]\Bigr)+\hskip 3truecm $$
$$\hskip 5.2truecm \Bigl({n(\Omega_0)\over \hbar}+1\Bigr)\Bigl([\hat a\hat\rho,\hat a^+]+
[\hat a,\hat\rho\hat a^+]\Bigr)\biggr)\hskip 2truecm (3.12)$$
$$\hskip 2truecm \Delta(\Omega_0)=\int\limits_0^{+\infty}d\omega\biggl({1\over{\omega
-\Omega_0}}+{1\over{\omega+\Omega_0}}\biggr){\vert\epsilon(\omega)\vert}^2
\sigma(\omega)\hskip 2.6truecm (3.13)$$
As already mentioned, (3.13) is identical with the well-known kinetic 
equation used in quantum optics and derived by making the rotating-wave
approximation ([8] eq.(6.2.59), [9] eq.(3b.7) p.118, [11] eq.(3.1), [36] eq.
(5.1.59)).
On the other hand (3.13) is of the form (5.8) of paper I, with the following
identifications:
$$\hskip 2truecm \tilde h_{\alpha\beta}(\omega)\leftrightarrow\delta_{n,m}
\quad ,\quad n,\ m=1,\ 2\hskip 2truecm $$
$$\hskip 1truecm \hat B_1 \leftrightarrow\hbar\gamma\sqrt{{n(\Omega_0)\over
\hbar}}\ \ \hat a^+\ \ \ \ \ 
\hat B_2 \leftrightarrow\hbar\gamma\sqrt{{n(\Omega_0)\over
\hbar}+1}\ \ \hat a\hskip 3.2truecm (3.14a)$$
$$\hskip 4.5truecm \gamma^2=\pi{\vert\epsilon(\Omega_0)\vert}^2\sigma(\Omega_0)\hskip 5.5truecm (3.14b)$$
Therefore its phase-space representation via {\it any involutive} generalized 
Wigner transformation (cf.(3.16) of paper I) follows from (5.17) of paper I.
From (3.2) we find the {\it Wigner transforms} of $\hat B_1, \hat B_2$
$$\hskip 1truecm B_1\equiv\gamma\sqrt{{n(\Omega_0)\over 2}}\Bigl(\sqrt
{\Omega_0}q-{ip\over\sqrt{\Omega_0}}\Bigr)\quad ,\quad B_2\equiv\gamma\sqrt{{n(\Omega_0)+\hbar
\over 2}}\Bigl(\sqrt{\Omega_0}q+{ip\over\sqrt{\Omega_0}}\Bigr)$$
and consequently their Fourier transforms (paper I, section2)
$$\tilde B_1(\eta,\xi)=2\pi i\sqrt{{n(\Omega_0)\over 2}}\gamma\Bigl(\sqrt{\Omega_0}\delta(\xi)
\delta '(\eta)-{i\over\sqrt{\Omega_0}}\delta '(\xi)\delta(\eta)\Bigr)$$
$$\tilde B_2(\eta ',\xi ')=2\pi i\sqrt{{n(\Omega_0)+\hbar\over 2}}\gamma\Bigl(\sqrt{\Omega_0}\delta(\xi ')
\delta '(\eta ')+{i\over\sqrt{\Omega_0}}\delta '(\xi ')\delta(\eta ')\Bigr)$$
Therefore (3.14) above and (5.18) of paper I finally give
$$B(\sigma ',\sigma)=-\Bigl(\Omega_0D_1\delta(\xi)\delta(\xi ')\delta '(\eta)\delta '(\eta ')+
{D_1\over \Omega_0}\delta '(\xi)\delta '(\xi ')\delta(\eta)\delta(\eta ')\Bigr)$$
$$\hskip 2truecm +iD_2\Bigl(\delta(\xi)\delta '(\xi ')\delta '(\eta)\delta(\eta ')-\delta '(\xi)\delta(\xi ')\delta(\eta)\delta '(\eta ')\Bigr)\hskip 2.8truecm (3.15)$$
$$\hskip 2truecm D_1=2\pi^2\gamma^2(2n(\Omega_0)+\hbar)\quad ,\quad D_2=2\pi^2\gamma^2\hbar\hskip 3.8truecm (3.16)$$
Substitution of (3.15) in (5.17) of paper I gives after a straightforward
calculation and with the aid of (2.12) of paper I, the phase-space 
representation of (3.12):
$${\partial\rho\over\partial t}={1\over i\hbar}\Bigl(1-\lambda^2{\Delta(\Omega_0)
\over\Omega_0}\Bigr)[H_\Sigma ,\rho]-\hskip 8.9truecm $$
$$-{2\lambda^2\over({2\pi})^3\hbar^2}\Re\biggl(-\Omega_0D_1\int d\sigma
e^{i\sigma z}\tilde\rho (\sigma)\Bigl(\mu^2\xi^2+i\mu\xi q-\mu\xi{\partial\chi
(\sigma)\over\partial\eta}\Bigr)\hskip 3truecm $$
$$+{D_1\over\Omega_0}\int d\sigma e^{i\sigma z}\tilde\rho (\sigma)\Bigl(-\mu^2
\eta^2+i\mu\eta p-\mu\eta{\partial\chi(\sigma)\over\partial\xi}\Bigr)$$
$$\hskip 1truecm +iD_2\int d\sigma e^{i\sigma z}\tilde\rho (\sigma)\Bigl(2\mu+i\mu(q\eta+
p\xi)-\mu\eta{\partial\chi(\sigma)\over\partial\eta}-\mu\xi{\partial\chi(\sigma)\over\partial\xi}\Bigr)\biggr)\hskip 1truecm (3.17)$$
where $\mu={i\hbar\over 2}$, [\ ,\ ] is the bracket with respect to the $*_\Omega$-
product, $H_\Sigma$ is the classical Hamiltonian of the oscillator and the
kernel of the generalized Wigner transformation is 
$$\hskip 2truecm \Omega(\sigma)=e^{\chi(\sigma)}\hskip 9.8truecm (3.18)$$
since $\Omega$ is an entire (analytic) function without zeros (cf. paper I,
section 2). Eq.(3.18) follows then from the Weierstrass factor theorem for
several variables (see e.g. [27]). Since the generalized Wigner transformation
is assumed involutive, so that 
$$\Omega(\sigma)=\Omega^*(-\sigma)\ \ \Leftrightarrow\ \ \chi(-\sigma)=\chi^*(\sigma)$$
(cf. (3.16) of paper I), we get 
$$\tilde\rho(-\sigma){\partial\chi(-\sigma)\over\partial(- \eta)}=-{\Bigl
(\tilde\rho(\sigma) {\partial\chi(\sigma)\over\partial\eta}\Bigr)}^*$$
that is the Fourier transform of $\tilde\rho(\sigma) {\partial\chi(\sigma)\over\partial\eta}$
(and for that matter, of $\tilde\rho(\sigma) {\partial\chi(\sigma)\over\partial\xi }$)
is an imaginary function hence it does not contribute to (3.17). The same is
true for the second term in the first two integrals. Using this, simple 
reductions transform (3.17) with the aid of (3.16), to
$${\partial\rho\over\partial t}={1\over i\hbar}\Bigl(1-\lambda^2{\Delta(\Omega_0)
\over\Omega_0}\Bigr)[H_\Sigma,\ \rho]+\hskip 9truecm $$
$$+{\lambda^2\gamma^2\over 2}\biggl[{\partial\over\partial p}\Bigl(\Omega_0(n(\Omega_0)
+{\hbar\over 2}){\partial\rho\over\partial p}+p\rho+p{\Psi\over2\pi}\odot\rho
\Bigr)+\hskip 3.6truecm $$
$$\hskip 2.5truecm +{\partial\over\partial q}\Bigl({1\over\Omega_0}\Bigl(n(\Omega_0)+
{\hbar\over 2}\Bigr){\partial\rho\over\partial q}+q\rho +q{\Psi\over 2\pi}
\odot\rho\Bigr)\biggl]\hskip 3.3truecm (3.19)$$
where
$$\hskip 2truecm \Psi(z)={1\over 2\pi}\int e^{i\sigma z}\chi(\sigma)d\sigma\hskip 7.3truecm (3.20)$$
and $\odot$ is the convolution product. This is the desired phase-space
equation for the quantum oscillator, obtained via {\it any} involutive
generalized Wigner transformation. It is now trivial to obtain the classical
limit of (3.19) (or equivalently of (3.12)), since all quantities, except 
$\Psi$ are $\hbar$-independent. However assuming that ${1\over i\hbar}[\ ,\ ]$
is a deformation of the Poisson bracket, proposition 3.2 of paper I and (3.18)
imply that if $(\hat q,\hat p)$ is mapped to $(q,p)$ then
$\lim\limits_{\hbar\to 0^+}\Psi(z)=0$, hence if $\lim\limits_{\hbar\to 0^+}\rho=
\rho_0$, then 
$${\partial\rho_0\over\partial t}=\Bigl(1-\lambda^2{\Delta(\Omega_0)\over
\Omega_0}\Bigr)\{H_\Sigma,\rho_0\}+\hskip 9truecm $$
$$\hskip 0.5truecm +{\lambda^2\gamma^2\over 2}\biggl({\partial\over\partial p}\Bigl(\Omega_0
n(\Omega_0){\partial\rho_0\over\partial p}+p\rho_0\Bigr)+{\partial\over\partial 
q}\Bigl({1\over\Omega_0}n(\Omega_0){\partial\rho_0\over\partial q}+q\rho_0
\Bigr)\biggr)\hskip 2truecm  (3.21)$$
Taking account of (3.9b), (3.14d) 
and the remarks at the end of Appendix 1
we see that (3.21) with $\sigma=1$ {\it is identical with
the classical equation (2.26)} since $u^2(\omega), \omega > 0$, corresponds to
$4\epsilon(\omega)\omega\Omega_0$ as mentioned at the beginning of this
section. When $\sigma\neq1$ the comparison should be made with the kinetic
equation following by using the Hamiltonian (2.8) instead of (2.11) in section
2. However in view of the comments made at the end of Appendix 1, once again
the two equations turn to be identical.
\vskip 2truecm
\leftline {\bf 4. KINETIC EQUATION FOR MORE GENERAL HARMONIC}
\leftline {\bf\hskip 0.6truecm OSCILLATOR MODELS}
\bigskip
In the previous section we obtained the kinetic equation for a quantum harmonic
oscillator, weakly coupled to an equilibrium bath of other oscillators, eq(3.12)
within the context of the general formalism of [5]. Moreover its phase-space
repsesentation via an arbitrary involutive generalized Wigner transformation was
obtained, eq.(3.19), from which the uniqueness of its classical limit,eq.(3.21),
follows. Direct application of (3.2) to (3.12) shows that the latter is a 
special case of the following more general equation
$${\partial\hat\rho\over\partial t}=-{i\over\hbar}[\hat H_\Sigma,\hat \rho]-
{i\over\hbar}(\Lambda+\kappa )[\hat q,[\hat\rho,\hat p]_+]+{i\over\hbar}(\Lambda
-\kappa)[\hat p,[\hat\rho,\hat q]_+]$$
$$\hskip 3truecm -{D_1\over\hbar^2}[\hat q,[\hat q,\hat \rho]]-{D_2\over\hbar^2}[\hat p,[\hat p,
\hat\rho]]+{D\over\hbar^2}([\hat q,[\hat p,\hat\rho]]+[\hat p,[\hat q,\hat \rho]])
\hskip 1truecm (4.1)$$
with $\hat H_\Sigma$ given by (3.1).
\par
This equation has been considered in the litterature as a master equation in 
quantum optics, describing an electromagnetic field mode interacting with an
equilibrium bath of bosons. It also includes many other master equations used 
to study open systems in heavy ion collisions ([13] section 3,[14] section 1,2
and references therein). As it will be seen, it is of the Lindblad type. In 
fact, it follows from it (eq.$(5.10'')$) of paper I) assuming that as functions
of $\hat q,\hat p$, its non dissipative part is quadratic and the operators
appearing in the dissipative part, are linear ([13] section 3, eq.(3.6)).
\par
In [14] its phase-space representation by means of Wigner's transformation or
its generalizations corresponding to antinormal and normal ordering of 
operators (Glauber repsesentation) have been given, showing that it is an 
equation of the Fokker-Planck type, i.e. with nonegative-definite 2nd order
term, provided that 
$$\hskip 2truecm D_1,D_2 > 0\quad\  D_1D_2-D^2\geq{\hbar^2\Lambda^2\over 4}\hskip 6truecm (4.2)$$
The formalism of paper I allows us to generalize
this result obtaining the phase-space analogue of (4.1) for {\it any involutive
generalized Wigner transformation}. The classical limit of (4.1) is then a
simple matter, and as for the special case of section 3, it turns out to be
unique.
\par
Rearranging the second term on the r.h.s. of (4.1), we get 
$${\partial\hat\rho\over\partial t}=-{i\over\hbar}[\hat H_\Sigma+\kappa[\hat q,\hat p]_+,\hat\rho]-\hskip 10truecm  $$
$$\hskip 2truecm -{1\over\hbar^2}(D_1[\hat q,[\hat q,\hat\rho]]+D_2[\hat p,[\hat p,\hat\rho]]-
D([\hat q,[\hat p,\hat\rho]]+[\hat p,[\hat q,\hat\rho]])$$
$$\hskip 2truecm -{i\Lambda\over\hbar}\biggl([\hat q,[\hat p,\hat\rho]_+]-[\hat p,[\hat q,\hat
\rho ]_+]\biggr)\hskip 6.5truecm (4.1')$$
Comparison with (5.5) of paper I shows that $(4.1')$ is of the same form with
the identifications
$${\lambda^2\over 2}\tilde h_{\alpha\beta}^q=\left(\matrix{D_1 &-D\cr -D &D_2
\cr}\right),\ \ \ \ \ {\lambda^2\over 2}\tilde g_{\alpha\beta}^q=\left(\matrix{0
&\Lambda\cr -\Lambda &0\cr}\right)$$
$$\hat V_1=\hat V_1^+=\hat q\quad,\quad V_2=\hat V_2^+=\hat p$$
Therefore it can be put in the form (5.8) of paper I, hence it is of the 
Lindblad type, (eq.$(5.10'')$ of paper I) with $\hat V_1,\hat V_2$ as above and
$$\hskip 2truecm {\lambda^2\over 2\hbar^2}\tilde h_{\alpha\beta}={1\over\hbar^2}
\left(\matrix {D_1 & {i\hbar\Lambda\over 2}-D \cr -{i\hbar\Lambda\over 2}-D &
D_2 \cr}\right) \hskip 4.8truecm (4.3)$$
{\it provided the latter is nonegative-definite, from which (4.2) follows).}
Hence its phase-space representation is given by (5.17) of paper I with
$$B_1\equiv V_1\equiv \Omega^{-1}_w(\hat V_1)=q,\quad B_2\equiv V_2\equiv 
\Omega^{-1}_w(\hat V_2)=p$$
and $\Omega^{-1}_w$ indicating the Wigner transformation (c.f. section 2.1 of
paper I). Consequently their Fourier transforms are
$$\hskip 2truecm \tilde B_1(\eta',\xi')=2\pi i\delta'(\eta')\delta(\xi')\ \ \ \ \  
\tilde B_2(\eta,\xi)=2\pi i\delta(\eta)\delta'(\xi)\hskip 2.5truecm (4.4)$$
so that with the usual notation $\sigma=(\eta,\xi), z=(q,p)$ etc, (4.3), (4.4)
substituted in (5.18) of paper I, give
$${\lambda^2\over 2}B(\sigma,\sigma')\equiv\sum_{n,m=1}^2\tilde V_n^*(\sigma')
\tilde h_{nm}\tilde V_m(\sigma)=$$
$$=-(2\pi)^2\biggl(D_1\delta'(\eta)\delta'(\eta')\delta(\xi)\delta(\xi')+D_2
\delta(\eta)\delta(\eta')\delta'(\xi)\delta'(\xi')+$$
$$\hskip 1truecm +({i\hbar\Lambda\over 2}-D)\delta'(\eta')\delta(\eta)\delta(\xi')\delta'
(\xi)-({i\hbar\Lambda\over 2}+D)\delta'(\eta)\delta(\eta')\delta(\xi)
\delta'(\xi')\biggr)\hskip 1truecm (4.5)$$
Substituting (4.5) to (5.17) of paper I gives the desired kinetic equation in 
phase-space. The calculations are tedious but not difficult in principle and
are given in Appendix 2. The result is
$${\partial \rho\over\partial t}=D_1{\partial^2\rho\over\partial p^2}+D_2
{\partial^2\rho\over\partial q^2}+2D{\partial^2\rho\over\partial p\partial q}+\hskip 8.5truecm $$
$$+{\partial\over\partial p}[(\Omega_0^2q+(\Lambda+\kappa)p)(\rho+{\Psi\over 
2\pi} \odot\rho)]+{\partial\over\partial q}[(-{p\over m}+(\Lambda-\kappa)q)
(\rho+{\Psi\over 2\pi}\odot\rho)]\ (4.6)$$
where as in the previous section $\Psi$ is given by (3.18), (3.20). Therefore
it is clear that if the generalized Moyal bracket defined by the generalized
Wigner transformation is a deformation of the Poisson bracket, and
$(\hat q,\hat p)$ are mapped to $(q,p)$ then, as in the previous section 
$\lim\limits_{\hbar\rightarrow 0}\Psi(q,p)=0$ and (4.6) has the {\it unique} classical
limit $(\lim\limits_{\hbar\rightarrow 0}\rho\equiv \rho_0)$
$${\partial\rho_0\over\partial t}=\{H_\Sigma,\rho_0\}+\hskip 12truecm $$
$$+{\partial\over\partial p}(D_1{\partial\rho\over\partial p}+D{\partial\rho\over
\partial q}+(\Lambda+\kappa)p\rho)+{\partial\over\partial q} (D_2{\partial\rho
\over\partial q}+D{\partial\rho\over\partial p}+(\Lambda-\kappa)q\rho)
\hskip 2truecm (4.7)$$
where $H_\Sigma$ is the classical hamiltonian for the  oscillator, (2.4a)
\par
To recover known results, we explicitly calculate the convolution terms in (4.6)
for
$$\hskip 2truecm \Omega(\eta,\xi)=e^{a{\hbar\over 4}(\Omega_0^2\xi^2+{\eta^2\over\Omega_0})}
\ \ \ \ \ a\in{\cal R}\hskip 6.4truecm (4.8)$$
The following cases are included in (4.8) ([28])
$$a=0\quad:\quad Weyl\ ordering\hskip 6.05truecm$$
$$a=-1\ :\quad normal\ ordering\ (Glauber\ representation)$$
$$a=1\quad:\quad antinormal\ ordering\hskip 4.77truecm $$
From (4.8), (3.20) we find
$$\Psi(q,p)=-{2\pi\lambda\hbar\over 4\Omega_0}(\delta(p)\delta''(q)+\Omega_0^2
\delta(q)\delta''(p))$$
Since $x\delta''(x)=-2\delta'(x),x\delta(x)=0$ a direct calculation
gives
$${\partial\over\partial p}[(\Omega_0^2q+(\Lambda+\kappa)p)({\Psi\over 2\pi}
\odot\rho)]+{\partial\over\partial q}[(-{p\over m}+(\Lambda-\kappa)q)({\Psi
\over 2\pi} \odot\rho)]=$$
$$=-{\hbar a\over 2}\Omega_0(\Lambda+\kappa){\partial^2\rho\over\partial p^2}
-{\hbar a\over 2\Omega_0}(\Lambda-\kappa){\partial^2\rho\over\partial q^2}$$
Substituting this in (4.6), we finally get 
$${\partial\rho\over\partial t}=(D_1-a\hbar{(\Lambda+\kappa)\over 2}\Omega_0)
{\partial^2\rho\over\partial p^2}+(D_2-a\hbar{(\Lambda-\kappa)\over 2\Omega_0})
{\partial^2\rho\over\partial q^2}+2D{\partial^2\rho\over\partial q\partial p}$$
$$\hskip 2truecm +{\partial\over\partial p}[(\Omega_0^2q+(\Lambda+\kappa)p)\rho]+
{\partial\over\partial q}[(-{p\over m}+(\Lambda-\kappa)q)\rho]\hskip 2.5truecm (4.9)$$
Introducing the variables
$$x_1=\sqrt{{\Omega_0\over 2\hbar}}q\quad\quad x_2={p\over\sqrt{2\hbar\Omega_0}}$$
So that $x=x_1+ix_2$ are the eigenvalues of $\hat a$ corresponding to 
coherent states (e.g. [8] p.107-108), (4.9) becomes
$${\partial \rho\over\partial t}=\biggl({D_1\over 2h\Omega_0}-a{(\Lambda+\kappa)\over
4}\biggr){\partial^2\rho\over\partial x_2^2}+\biggl({D_2\Omega_0\over 2\hbar}-a{(
\Lambda -\kappa)\over 4}\biggr){\partial^2\rho\over\partial x_1^2}+{D\over\hbar}
{\partial^2\rho\over\partial x_1\partial x_2}$$
$$\hskip 2truecm {\partial\over\partial x_1}[((\Lambda-\kappa) x_1-\omega x_2)\rho]+
{\partial\over\partial x_2}[((\Lambda+\kappa) x_2+\omega x_1)\rho]\hskip 2truecm (4.9')$$
For $a=0,\pm 1$ this is identical with eq. (4.5) and Table III of [14], see 
also [13] eq.(5.19) for the case $a=0$
\bigskip
\bigskip
In the present paper the general formalism of paper I for the phase-space 
representation of quantum kinetic equations, has been applied to models of a
harmonic oscillator damped by its interaction with an equilibrium bath of
other harmonic oscillators. These equations are of the Lindblad type, but it
should be emphasized that {\it the methods of paper I allow for a phase-space
representation of any kinetic equation the evolution operator of which is 
formed by algebraic operations between quantum operators}. Other applications,
as well as some more fundamental problems of kinetic theory, already briefly 
discussed in paper I section 1 and [29] section 4, will be examined in the third
paper of this series.
\vskip 2truecm
\centerline{\bf APPENDIX 1}
\bigskip
Here we derive eqs (2.22), (2.23) by using simple functional integration
(see e.g. [24]):
By (2.18), $H_R$ is quadratic in $\phi ,\pi$ hence by the formula
$${\delta\over\delta\phi(\omega)}\Bigl(e^{-\beta H_R}\Bigr)=
-\beta{\delta H_R\over\delta\phi(\omega)}\ e^{-\beta H_R}$$
and similarly for $\pi(\omega)$ we get
$$\int\delta\phi\delta\pi\ \phi(\omega)\phi(\omega')\rho_R=-
\int\delta\phi\delta\pi\ {\phi(\omega)\over\beta\omega'^2}
{\delta\rho_R\over\delta\phi(\omega')}=$$
$$={1\over\beta\omega'^2}\int\delta\phi\delta\pi\ {\delta\phi(\omega)\over
\delta\phi(\omega')}\rho_R={1\over\beta\omega'^2}\delta(\omega-\omega')$$
where we have used integration by parts and have taken into
account that
$$\int\delta\phi\delta\pi\ {\delta\over\delta\phi}\Bigl(\phi^2\rho_R\Bigr)=0$$
Similar calculations give
$$\int\delta\phi\delta\pi\ \pi(\omega)\pi(\omega')\rho_R={1\over\beta}\delta(\omega-\omega')$$
$$\int\delta\phi\delta\pi\ \pi(\omega)\phi(\omega')\rho_R=0$$
$$\int\delta\phi\delta\pi\ \phi(\omega)\phi(\omega')={1\over\beta\omega^2}\delta(\omega-\omega')$$
Using this and (2.14b) in (2.17a) we find 
$$\hskip 2truecm h(s)=\int d\omega{u^2(\omega)\over\beta\omega^2}\cos{\omega s}\hskip 6.5truecm (A.1.1)$$
Similarly, using
$${\delta\phi(\omega)\over\delta\phi(\omega')}={\delta\pi(\omega)\over\delta\pi(\omega')}
=\delta(\omega-\omega'),\quad\quad\quad {\delta\phi(\omega)\over\delta\pi(\omega')}=0$$
and that $\int \delta\phi\delta\pi\ \rho_R=1$ , we find from (2.17b) that
$$\hskip 2truecm g(s)=\int d\omega{u^2(\omega)\over\omega}\sin{\omega s}\hskip 6.6truecm (A.1.2)$$
Using the well-known formulas
$$\int\limits_{-\infty}^{+\infty}ds\ e^{ias}=2\pi\delta(a),
\ \ \ \ \ \int\limits_0^{+\infty}ds\ e^{ias}=\pi\delta(a)+{i\over a}$$
we get
$$\hskip 1truecm \int\limits_{-\infty}^{+\infty}ds\ e^{ias}\cos{\omega s}=\pi\Bigl(\delta(a-\omega)+\delta(a+\omega)\Bigr)\hskip 4.3truecm (A.1.3a)$$
$$\hskip 1truecm \int\limits_{-\infty}^{+\infty}ds\ e^{ias}\sin{\omega s}=\pi\Bigl(\delta(a-\omega)-\delta(a+\omega)\Bigr)\hskip 4.4truecm (A.1.3b)$$
$$\hskip 1truecm \int\limits_0^{+\infty}ds\ e^{ias}\cos{\omega s}={\pi\over 2}\Bigl(\delta(a-\omega)+\delta(a+\omega)\Bigr)+{ia\over a^2-\omega^2}\hskip 2.8truecm (A.1.3c)$$
$$\hskip 1truecm \int\limits_0^{+\infty}ds\ e^{ias}\sin{\omega s}={i\pi\over 2}\Bigl(\delta(a-\omega)-\delta(a+\omega)\Bigr)-{\omega\over a^2-\omega^2}\hskip 2.7truecm (A.1.3d)$$
Substituting (A.1.1) in (2.17a) and using (A.1.3a) we get 
$$\hskip 2truecm \tilde h(a)=\pi\int \Bigl(\delta(a-\omega)+\delta(a+\omega)\Bigr)
{u^2(\omega)\over\beta\omega^2}d\omega\hskip 4truecm (A.1.4)$$
which gives the first of (2.22). The expressions for $\tilde g,\bar h,\bar g$
in (2.22), (2.23$'$) are similarly obtained. We may notice here that if
instead of the Hamiltonian (2.11) we consider (2.8), then nothing changes in
the preceding calculations as well as those of section 2, except that the
integration in (A.1.1), (A.1.2) are with respect to $\theta$. Hence in (A.1.3)
$\omega$ is replaced by $\omega(\theta)$ and consequently by a change of
variables $\theta\tini\omega(\theta)$, (A.1.4) gives
$$\hskip 2truecm \tilde h(a)={2\pi{\Bigl(u(\theta(a))\Bigr)}^2\over a^2}\sigma(a)
\quad\quad ,\quad\quad \sigma(\omega)={d\theta\over d\omega}\hskip 3truecm $$
provided $\omega(\theta)$ is an invertible, even function of $\theta$ (notice
that because of (2.10), $\omega(\theta)$ is necessarily either even or odd).
It is now clear that in the kinetic equation (2.26) (and in (2.25) as well),
$u^2$ is replaced by $u^2\sigma$, which has already been remarked in section
2.
\par
If the spectral variable $\omega$ in (2.11) is nonegative then $u(\omega)$ can
be extended by putting $u(\omega)=0$, for $\omega\leq 0$ and all calculations
remain unaltered except that 
\smallskip
\ind (i) in (A.1.4) (and the corresponding equation for $\tilde g$), the one 
$\delta$-function does not contribute

\ind (ii) by (2.22), $\tilde h(\omega_{-1})=\tilde g(\omega_{-1})=0$, hence
the 2nd term of (2.21) does not contribute to (2.15).
\smallskip
\ind Then it is readily checked that these changes imply a factor $1\over 4$
on the r.h.s. of (2.24). The case $\omega >0$ is used in section 3 to show
explicitly that (2.26) thus modified, is the classical limit of the kinetic
equation for the corresponding quantum system.
\vskip 2truecm
\centerline{\bf APPENDIX 2}
\bigskip
Here we derive (4.6). For the dissipative part we substitute (4.5) to eq(5.17)
of paper I. Specifically we compute
$$-{2\lambda^2\over {(2\pi)}^3\hbar^2}\Re\int d\sigma d\sigma' d\sigma''
{e^{i(\sigma+\sigma'+\sigma'')}\over\Omega(\sigma+\sigma'+\sigma'')}
B(\sigma',\sigma)\hskip 6truecm $$
$$\tilde\rho_w(\sigma'')e^{\mu(\sigma'\land\sigma'')}
\sinh{\mu}\Bigl(\sigma\land(\sigma'+\sigma'')\Bigr)$$
where $\tilde\rho_w$ is the Fourier transform of the Wigner transform of
$\hat\rho$, $\sigma=(\eta,\xi)$, $\sigma\land\sigma'=\eta\xi'-\eta'\xi$ (see
section 2 of paper I for the notation). Substitution of $B(\sigma',\sigma)$
from (4.5) and straightforward integrations of the $\delta$-functions give
$${4\over 2\pi\hbar^2}\Re\int d\sigma e^{i\sigma z}\tilde\rho(\sigma)\biggl[
D_1\Bigl(\mu^2\xi^2+i\mu\xi q-\mu\xi\partial_\eta\chi(\sigma)\Bigr)+\hskip 5truecm $$
$$+D_2\Bigl(\mu^2\eta^2-i\mu\eta p+\mu\eta\partial_\eta\chi(\sigma)\Bigr)
-\Bigl({i\hbar\Lambda\over 2}-D\Bigr)\Bigl(\mu+i\mu q\eta+\mu^2\eta\xi-
\mu\eta\partial_\eta\chi(\sigma)\Bigr)$$
$$-\Bigl({i\hbar\Lambda\over 2}+D\Bigr)\Bigl(\mu+i\mu p\xi-\mu^2\eta\xi-
\mu\xi\partial_\xi\chi(\sigma)\Bigr)\biggr]\hskip 6.5truecm $$
where we used (2.12) of paper I and (3.18). It is not difficult to see that 
this is $\Bigl(\mu={i\hbar\over 2}\Bigr)$
$$D_1{\partial^2\rho\over\partial p^2}+D_2{\partial^2\rho\over\partial q^2}
+2D{\partial^2\rho\over\partial q\partial p}+\Lambda{\partial\over\partial p}
\Bigl[p\Bigl(\rho+{\Psi\over 2\pi}\odot\rho\Bigr)\Bigr]+
\Lambda{\partial\over\partial q}\Bigl[q\Bigl(\rho+{\Psi\over 2\pi}\odot\rho\Bigr)\Bigr]
\hskip 0.4truecm (A.2.1)$$
For the nondissipative term of (4.1$'$) we proceed as follows: Since
$\hat H_\Sigma$ is given by (3.1), (3.2) its Wigner transform is the classical
Hamiltonian
\ind $H_\Sigma={p^2\over 2}+\Omega^2_0q^2$ hence its Fourier transform is
$$\tilde H_\Sigma=-2\pi\Bigl({\delta(\eta)\delta''(\xi)\over 2}+\Omega^2_0
\delta''(\eta)\delta(\xi)\Bigr)\hskip 3truecm $$
Substituting this in (5.12b) of paper I we get after some simple reductions
and using (2.12) of paper I:
$$[H_\Sigma,\rho]={1\over 2\pi}\biggl(-{1\over 2}\int d\sigma e^{i\sigma z}
\tilde\rho(\sigma)\Bigl(4i\mu\eta p-4\mu\eta\partial_\xi\chi(\sigma)\Bigr)$$
$$\hskip 2.5truecm +{\Omega^2_0\over 2}\int d\sigma e^{i\sigma z}\tilde\rho(\sigma)\Bigl(
4i\mu\xi q-4\mu\xi\partial_\eta\chi(\sigma)\Bigr)\biggr)$$
This is easily transformed to
$$\hskip 1truecm -{i\over\hbar}[H_\Sigma,\rho]=\{H_\Sigma,\rho\}+\Omega^2{\partial\over
\partial p}\Bigl(q{\Psi\over 2\pi}\odot\rho\Bigr)-{\partial\over\partial q}
\Bigl(p{\Psi\over 2\pi}\odot\rho\Bigr)\hskip 2truecm (A.2.2)$$
Finally, since $(\hat q,\hat p)$ is mapped to $(q,p)$ by the Wigner 
transformation, a similar calculation using (5.15b) of paper I yields
$$-{i\over\hbar}\kappa\Bigl[[q,p]_+,\rho\Bigr]=\kappa\biggl(p{\partial\rho
\over\partial p}-q{\partial\rho\over\partial q}-{\partial\over\partial q}
\Bigl(q{\Psi\over 2\pi}\odot\rho)\Bigr)+{\partial\over\partial p}
\Bigl(p{\Psi\over 2\pi}\odot\rho)\Bigr)\biggr)\hskip 1truecm (A.2.3)$$
Combining (A.2.1-3) gives (4.6).
\vskip 2truecm
\centerline{\bf APPENDIX 3}
\bigskip
In sections 2, 3 we derive kinetic equations for a harmonic oscillator, 
weakly-coupled to a harmonic chain, by applying the general formal theory of
open systems $\Sigma$ weakly coupled to equilibrium baths R, developped in 
[5] (see also paper I, section 4.2). The starting point of this formalism
is the solution of the Liouville (or von Neumann) equation for the total 
system, for positive times. This implies that the corresponding resolvent
operator is analytic on the complex upper half-plane and has a cut along 
(part of) the real axis, which contains its necessarily existing continuous 
spectrum. Assuming that the restriction of this resolvent to the state
space of the open system admits a meromorphic analytic continuation 
through the cut in the lower half-plane, it can be shown that the corresponding
exponentially decaying contributions to the solution of the exact dynamics
satisfy a Markovian equation. Moreover, its generator can be given a 
representation in the subspace of the open system under certain assumptions
which will not be discussed here (cf. [5] section 2). This generator, $\Phi$
say, has been computed to second order in the coupling parameter of the 
open system with the bath ([5] section 3) and it has been shown to be 
identical to that proposed by Davies ([3] eq.(2.17). See also [32]). It is used
in this paper for the particular system studied. Specifically if the
Liouville operator $L$ of the total system is splitted as (see (4.6) of paper
I)
$$L=L_\Sigma+L_R+\lambda L_I\equiv L_0+\lambda L_I$$
and $P$ is the projection onto the subspace of the open system (see (4.7)
of paper I), then to second order in the coupling parameter $\lambda,\ \Phi$
is ([5] eq.(3.11$'$))
$$\Phi=-iPLP-\lambda^2\lim_{T\tini +\infty}{1\over 2T}\int\limits_{-T}^Tdt
\int\limits_0^{+\infty}ds\ Pe^{iL_0(s+t)}L_IQe^{-iL_0s}QL_Ie^{-iL_0t}P$$
where $Q$ is the complementary projection to $P$. Here it has been assumed
that $P$ commutes with $L_0$, i.e. for $\lambda=0$, $\Sigma$ evolves 
independently of R which stays in equilibrium, and that $L_\Sigma$ has a point
spectrum.
\par
On the other hand, projecting Liouville's (or von Neumann's) equation to the
$P$ subspace, the so-called Generalized Master Equation (GME) results, which
under the same assumptions, lead to {\it a quite different generator}, $\Theta$
say, ([33] section 2, eq.(2.10) and [5] eq.(3.7$'$))
$$\Theta=-iPLP-\lambda^2\int\limits_0^{+\infty}ds\ PL_IQe^{-iL_0s}QL_Ie^{iL_0s}P$$
Notice that 
$$\Phi=\lim_{T\tini +\infty}{1\over 2T}\int\limits_{-T}^Tdt\ e^{iL_0t}\Theta e^{-iL_0t}$$
$\Theta$ is essentially the generator that lead to kinetic equations with
serious defects (nonconservation of positivity of the states and for that
matter, no H-theorem), like (2.27) for classical systems (see the discussion
in [5] section 5) and which call for additional assumptions, like the
rotating-wave approximation mentioned
in section 3, in order to get rid of these defects.
\par
For a separable interaction Hamiltonian, $\Phi$ takes the form (4.22$'$) of
[5], which for the model of section 2 gives (2.15), whereas for its quantum 
analogue it gives (3.11) (notice that $PL_IP=0$ in our case).
\vfill{\eject}
\centerline{\bf REFERENCES}
\bigskip
\ind 1) C. Tzanakis and A.P. Grecos, {\it "Generalized Moyal structures in phase space,

\ind \hskip 0.6truecm kinetic equations and their classical limit. I: General Formalism"} (Preprint 

\ind \hskip 0.6truecm 1996)
\smallskip
\ind 2) G. Lindblad, Comm. Math. Physics {\bf 48} (1976) 119
\smallskip
\ind 3) E. B. Davies, Comm. Math. Physics {\bf 39} (1974) 91
\smallskip
\ind 4) H. Spohn and J. L. Lebowitz, Adv. Chem. Physics {\bf 38} (1978) 109
\smallskip
\ind 5) A. P. Grecos and C. Tzanakis, Physica A {\bf 151} (1988) 61
\smallskip
\ind 6) C. Tzanakis, Ph. D. Thesis (Universit\'e Libre de Bruxelles 1987)
\smallskip
\ind 7) R. D\"umcke and H. Spohn Zeit. Physik B {\bf 34} (1979) 419
\smallskip
\ind 8) W. H. Louisell, {\it "Quantum statistical properties of radiation"} (Wiley-
Inter-

\ind \hskip 0.6truecm science, New York 1973)
\smallskip
\ind 9) F. Haake in {\it "Statistical Treatment of open system by Generalized
master} 

\ind \hskip 0.6truecm {\it equations"} Springer Tracks in Modern Physics Vol {\bf 66} (Springer, Berlin 1973)
\smallskip
\ind 10) H. Haken {\it "Light waves, photons, atoms"} (North-Holland, Amsterdam 

\ind \hskip 0.8truecm 1981)
\smallskip
\ind 11) G. S. Agarwal, Phys. Rev A {\bf 4} (1971) 739
\smallskip
\ind 12) G. S. Agarwal, {\it "Quantum statistical theories of spontaneous emission and} 

\ind \hskip 0.8truecm {\it their relation to other approaches"}, Springer Tracks in Modern Physics

\ind \hskip 0.8truecm Vol {\bf 70} (Springer, Berlin 1974)
\smallskip
\ind 13) A. Sandulescu and H. Scutaru, Ann. Physics {\bf 173} (1987) 277
\smallskip
\ind 14) A. Isar, W. Scheid and A. Sandulescu, J. Math. Physics {\bf 32} (1991)
2128
\smallskip
\ind 15) P. Ullersma, Physica {\bf 32} (1966) 27
\smallskip
\ind 16) P. Ullersma, Physica {\bf 32} (1966) 74
\smallskip
\ind 17) R. I. Cukier and P. Mazur, Physica {\bf 53} (1971) 157 eq.(1)
\smallskip
\ind 18) G. Ford, M. Kac and P. Mazur, J. Math. Phys. {\bf 6} (1965) 504, eq.(2)
\smallskip
\ind 19) D. Vitali and P. Grigolini, Phys. Rev. A {\bf 39} (1989) 1486 eq.(2.1)
\smallskip
\ind 20) K. Kobayashi and E. I. Takizawa, Mem. Faculty Engineering, Nagoya 

\ind \hskip 0.8truecm University {\bf 20} (1968) 144, eq.(2.1)
\smallskip
\ind 21) P. Ullersma, Physica {\bf 32} (1966) 56
\smallskip
\ind 22) H. Carmichael, {\it "An open system approach to quantum optics"}

\ind \hskip 0.8truecm Springer Lecture Notes in Physics m.18 (1993), p.13, 
eq (1.47) 
\smallskip
\ind 23) C. Tzanakis, {\it "Linear kinetic equations for finite classical systems: 

\ind \hskip 0.8truecm Explicit form and its mathematical and physical foundations"}, Preprint, 

\ind \hskip 0.8truecm University of Crete (1995)
\smallskip
\ind 24) M. H. Zaidi, Forschr. Phys. {\bf 31} (1983) 403 

\ind \hskip 0.8truecm M. J. Beran, {\it "Statistical continuum theories"} (Wiley-interscience, New 

\ind \hskip 0.8truecm York 1968)
\smallskip
\ind 25) G. Rosen, {\it "Formulation of classical and quantum dynamic theory"}
, (Academic press 1969)
\smallskip
\ind 26) P. Hatjimanolaki, Ph.D. Thesis (Universit\'e Libre de Bruxelles 1988)
\smallskip
\ind 27) K. Knopp, {\it "Theory of functions: Part II"}, (Dover, New York 1947), 
ch. I
\smallskip
\ind 28) G. V. Dunne, J. Phys. A {\bf 21} (1988) 2321
\smallskip
\ind 29) C. Tzanakis, Physica A {\bf 179} (1991) 531
\smallskip
\ind 30) S. A. Adelman, J. Chem. Physics {\bf 64} (1976) 124, eq (3.19)
\smallskip
\ind 31) N. G. van Kampen and I. Oppenheim, Physica {\bf 138A} (1986) 231 eq.(3.5)
\smallskip
\ind 32) E. B. Davies, Math. Annalen, {\bf 219} (1976) 147

\ind \hskip 0.7truecm E. B. Davies, Ann. Inst. H. Poincar\'e {\bf A28} (1978)
91
\smallskip
\ind 33) C. Tzanakis and A. P. Grecos, Physica A {\bf 149} (1988) 232
\smallskip
\ind 34) N. Corngold, Phys. Rev. {\bf A15} (1977) 2454

\ind \hskip 0.7truecm N. Corngold, Phys. Rev. {\bf A24} (1981) 656
\smallskip
\ind 35) E. B. Davies, Comm. Math. Physics {\bf 33} (1973) 171
\smallskip
\ind 36) C. W. Gardiner, {\it "Quantum Noise"}, (Springer Berlin 1991)
\end